\documentstyle[twocolumn,twoside,prb,floats,aps]{revtex}
\input{psfig}
\tighten
\begin{document}
\draft

\title{Parametric S-matrix fluctuations in quantum theory of
chaotic scattering}

\author{A. M. S. Mac\^edo}

\address{
Theoretical Physics, University of Oxford,
1 Keble Road, Oxford OX1 3NP, United Kingdom\\
{\rm (Submitted 8 april 1994)}\\
\parbox{14 cm}{\medskip\rm\indent
We study the effects of an arbitrary external perturbation in the statistical
properties of the S-matrix of quantum chaotic scattering systems in the limit
of isolated resonances. We derive, using supersymmetry, an exact
non-perturbative expression for the parameter dependent autocorrelator of two
S-matrix elements. Universality is obtained by appropriate rescaling of the
physical parameters. We propose this universal function as a new signature of
quantum chaos in open systems.}}
\maketitle
\narrowtext

Quantum chaotic scattering is the study of quantum transport in systems whose
underlying classical dynamics is chaotic. It has been recently the subject of
intense theoretical and experimental investigations \cite{chaos}. This growing
interest has been partially motivated by a large number of applications in many
branches of physics, such as microwave transmission through irregular shaped
cavities \cite{microwave}, ballistic transport in semiconductor nanostructures
\cite{mesoscopic} and resonant reaction theory in molecular and nuclear physics
\cite{nuclear}.

A generic set-up for a two-probe quantum chaotic scattering problem consists of
an interaction region of finite volume (the resonant cavity in a microwave
experiment or the ballistic microstructure in mesoscopic physics) connected to
two reservoirs by free propagation regions (wave guides for microwaves or
perfectly conducting leads for electron waves) wherein asymptotic scattering
channels can be defined. The principal role of the interaction region is to
provide a mechanism for "trapping" the incomming waves by irregular boundary
scattering thereby driving the system to a regime where the ray optic limit (or
classical dynamics) is dominated by classical chaos.

There are two main theoretical descriptions of this problem: the semiclassical
treatment \cite{chaos} and the stochastic approach \cite{stochastic,verb}. In
the semiclassical description, averages are calculated by representing the
S-matrix as a sum over all classical trajectories connecting two given
scattering channels \cite{gutz}. Predictions are quantitatively accurate only
for systems containing a large number of open channels. In the stochastic
approach the calculation of averages over actual parameters of the physical
system is replaced by averages over an ensemble of random matrices (the
Hamiltonian describing the interaction region or the S-matrix of the whole
scattering problem). This procedure, which relies on ergodicity, can only be
justified on time scales sufficiently large so that classical chaotic
scattering dominates and the incomming waves form long-lived resonances in the
interaction region.

The use of random matrix theory (RMT) in quantum chaos is now quite widespread
and has been largely motivated by the fact that it provides a natural framework
for obtaining quantum signatures of chaotic behavior. A striking example is the
Wigner-Dyson statistics \cite{WD} for level spacing distribution, which owing
to its remarkable robustness and universality can be considered the hallmark of
quantum chaos \cite{haake}.

In a very interesting recent series of papers \cite{sim1,sim2,sim3} Simons,
Altshuler, Lee and Szafer have extended RMT to describe the response of the
spectra of disordered and quantum chaotic systems to an external adiabatic
perturbation that causes the levels to disperse in disjoint manifolds
exhibiting many avoided crossings as the perturbation parameter varies. It has
been demonstrated that the $n$-point function for density of states
fluctuations becomes a universal function if the system dependent parameters
are removed by appropriate rescalings. More generally, their results can be
interpreted as a new signature of quantum chaos in closed systems.

For open systems, such as the quantum scattering set-up described above, the
injection and emission of waves through the contacts provides a mechanism for
level broadening and the above authors' analysis does not apply. In this case,
the search for new quantum signatures of chaotic behavior is still open and is
the main objective of the present work.

More precisely, we study the effects of an external adiabatic perturbation on
the fluctuation pattern of elements of the random S-matrix describing quantum
transport through a chaotic cavity weakly coupled to external reservoirs. We
demonstrate, by explicit calculation of the ensemble average using
supersymmetry, that the two-point autocorrelator of two  S-matrix elements at
different values of the external perturbation parameter, $U$, becomes a
universal expression if $U$ is measured in units of the mean square gradient of
the energy levels and if the decay width $\Gamma$ is measured in units of the
mean level spacing. We consider both, systems with and without time reversal
symmetry (T-symmetry).

The technique used in the present work permits, in fact, the complete solution
of the problem for an arbitrary number of channels and strength of the
couplings to the external reservoirs. The motivation to restrict our analysis
to the particular case of weak coupling is two-fold. First, the final
expression for the two-point function simplifies enormously and consequently
the physical understanding of each component becomes straightforward. Second,
this regime is highly non-perturbative and thus allows a deeper comprehension
of the limitations of a semiclassical treatment. The complete analysis of the
more general case will be discussed elsewhere \cite{macedo}.

A general two-probe quantum chaotic scattering problem can be explicitly set up
as follows. Let $|\psi^c_a(E)\rangle$ represent scattering eigenfunctions
inside the right ($c=R$) and left ($c=L$) free propagation regions, in which
$a=1,2,\dots,\Lambda$ labels the physical scattering channels. We denote the
completet set of orthonormal states characterizing the interaction region by
$|\mu \rangle$ and thus the Hamiltonian of the whole system, written in the
basis $\{|\psi^c_a(E)\rangle,|\mu \rangle \}$, is given by \cite{verb}
\begin{eqnarray}
{\cal H}_\beta&=&\sum_{a c} \int dE |\psi^c_a(E)\rangle E \langle\psi^c_a(E)| +
\sum_{\mu \nu}|\mu \rangle (H_\beta(U))^{}_{\mu \nu}  \langle \nu |
\nonumber \\
&&+\sum_{\mu a c}\int dE \{ |\mu \rangle W^c_{\mu a}
\langle \psi^c_a(E)|+ h.c.\},
\label {hamiltonian}
\end{eqnarray}
where $H_\beta(U)$ represents the projection of the full Hamiltonian onto the
interaction region. We assume that $H_\beta(U)$ can be written as \cite{sim3}
$H_\beta(U)=H_\beta(0)+U V_\beta$, where $H_\beta(0)$ is a member of the
Gaussian unitary ensemble (GUE) for systems without T-symmetry ($\beta=2$) and
belongs to the Gaussian orthogonal ensemble (GOE) for systems with T-symmetry
($\beta=1$), while $V_\beta$ is a fixed traceless Hamiltonian belonging to the
same ensemble as $H_\beta(0)$. This assumption has been shown in Ref.\
\onlinecite{sim3} to be sufficient to generate the same effective Lagrangian in
the nonlinear $\sigma$-model representation as the zero-mode approximation of
microscopic models, therefore it constitutes an accurate description of the
universal region. The third term in (\ref{hamiltonian}) represents the coupling
between the eigenstates of the interaction region and the scattering states in
the free propagation region.

For any finite number $N$ of bound states in the interaction region (we shall
later take $N \to \infty$ at the end of the calculation) the kernel of the
Lippmann-Schwinger equation is of finite rank, thus the S-matrix of the problem
can be calculated algebraically \cite{Weid} and after some staightforward
simplifications yields
\begin{equation}
S^{c c'}_{a b}(U)=\delta^{c c'}\delta_{a b}-2i\pi\sum_{\mu \nu}
W^c_{\mu a} (D^{-1})^{}_{\mu \nu}W^{c'}_{\nu b},
\label {S-matrix}
\end{equation}
in which
$$
D_{\mu \nu}\equiv (E+i0^+)\delta_{\mu \nu}-(H_\beta(U))^{}_{\mu \nu}
+i\pi \sum_{a c} W^c_{\mu a}W^c_{\nu a}.
$$
The absence of direct transitions between the physical channels enables us to
work in a representation that satisfies the requirement
$\sum_\mu W^c_{\mu a}W^{c'}_{\mu b}=N\delta^{c c'}\delta_{a b}x^c_a$. The weak
coupling regime is obtained simply by requiring $x^c_a \ll \Delta$, where
$\Delta$ is the mean level spacing. Using Eq. (\ref{S-matrix}) one can show
that the transmission probabilities defined as $T^c_a\equiv 1-|\langle S^{c
c}_{a a}\rangle |^2$ are given by $T^c_a\sim 4\pi^2 x^c_a/\Delta \ll 1$, which
physically corresponds to the limit of weakly overlapping levels, where
transport is dominated by isolated resonances.

The simplest two-point function that contains sufficient information about the
perturbation driven fluctuations in the elements of the S-matrix of the problem
is given by
\begin{equation}
\chi_\beta(U)=\sum_{a b c c'}\langle S^{c c'}_{a b}(\bar u)
{S^{c c'}_{a b}}^{*}(\bar u+U)\rangle,
\label{definition}
\end{equation}
where the angular brackets denote the usual ensemble average. We remark that a
correlation function of a form similar to this but taken as a function of
energy in absence of an external perturbation has been recently measured by
Doron, Smilanski and Frenkel \cite{microwave} in a microwave experiment. It is
also interesting to observe that for applications in mesoscopic physics the
average Landauer conductance is simply given by $\bar
G_\beta=(e^2/h)\chi_\beta(0)$.

We now state our results. Taking $N$ and $\Lambda$ to infinity such that $T^c_a
\to 0$, but with $T^c=\sum_a T^c_a$ remaining finite \cite{verb2},
$\chi_\beta(U)$ can then be calculated exactly using the conventional mapping
\cite{verb,ef83} of RMT onto the  zero-dimensional non-linear supersymmetric
$\sigma$-model. Integrating over the compact manifold of the massless
transverse modes by means of Efetov's parametrization \cite{ef83} of the
auxiliary supermatrix $Q$-fields, we find
\begin{equation}
\chi_\beta(U)={\pi \over \Delta} \Gamma T I_\beta({\pi \over \Delta} \Gamma,{U
\pi \over 2} (\beta C_0)^{1/2}),
\label {mainresult}
\end{equation}
where $T\equiv (1/T^R+1/T^L)^{-1}$ is the total transmission coefficient across
the interaction region, $\Gamma\equiv \Delta (T^R+T^L)/(2\pi)$ is the total
decay width for wave emission into the free propagation region and  $C_0$ is
the average gradient of level velocities as defined in Refs.\
\onlinecite{sim1,sim2,sim3}. We remark that Eq. (\ref{mainresult}) can also be
derived from a microscopic model of a particle diffusing in a disordered
potential by confining the saddle point Lagrangian to the lowest harmonic in
the spatial dependence of the composite supermatrix fields using the technique
of Refs. \onlinecite{ef83,sim1}. Finally, the function $I_\beta(x,y)$ can be
written as
$$
I_\beta(x,y)=\int_{(\beta)} d{\underline \lambda}\mu_\beta({\underline
\lambda}) f_\beta({\underline \lambda})
\exp(-x g_\beta({\underline \lambda})-y^2 f_\beta({\underline \lambda})),
$$
in which
$$
\int_{(1)} d{\underline\lambda}=\int_1^{\infty}d\lambda_1
\int_1^{\infty}d\lambda_2 \int_{-1}^{1} d\lambda_3 ,
$$
$$
\int_{(2)} d{\underline\lambda}=\int_1^{\infty}d\lambda_1 \int_{-1}^{1}
d\lambda_2,
$$
$$
\mu_1({\underline\lambda})={(1-\lambda_3^2) \over
(\lambda_1^2+\lambda_2^2+\lambda_3^2-2\lambda_1\lambda_2\lambda_3-1)^2},
$$
$$
f_1({\underline\lambda})=
2\lambda_1^2\lambda_2^2-\lambda_1^2-\lambda_2^2-\lambda_3^2+1,
$$
$$
\mu_2({\underline\lambda})=(\lambda_1-\lambda_2)^{-2},\qquad
f_2({\underline\lambda})=\lambda_1^2-\lambda_2^2,
$$
$$
g_1({\underline\lambda})=\lambda_1\lambda_2-\lambda_3, \qquad
g_2({\underline\lambda})=\lambda_1-\lambda_2.
$$
Observe that $\chi_\beta(U)$ becomes a universal function after the rescalings
\cite{sim1,sim2,sim3} $\Gamma \to \Delta \hat \Gamma$ and $U \to U_c
\hat U$, where $U_c=C_0^{-1/2}$. Eq. (\ref{mainresult}) is the complete
solution of the problem and can be interpreted as a new signature of quantum
chaos in open systems in the weak coupling regime.

The function $I_\beta(x,y)$ is displayed in Fig.\ \ref{f1} for $\beta=1$ and 2.
\begin{figure}
\centerline{\psfig{file=figure1.ps,width=8.0cm}}
\vspace{0.5cm}
\caption{$I_\beta(x,y)$ for $x=1$ as a function of $y$ for $\beta=1$ (GOE) and
$\beta=2$ (GUE) as indicated.
\label{f1}}
\end{figure}
Its typical Lorentzian-like tails reflects the long range nature of the
universal logarithmic eigenvalue repulsion of RMT. One can see two interesting
limits: (I) $U\gg U_c$ and (II) $U\ll U_c$.

(I) - In this case the levels decorrelate asymptotically and diagrammatic
perturbation theory applies. Alternatively, we can calculate $I_\beta(x,y)$
asymptotically for $y\gg1$ to find $I_\beta(x,y)\sim 1/y^2$. Thus
$\chi_\beta(U)$ acquires, after rescaling, the simple universal form
\begin{equation}
\hat \chi_\beta(U) \simeq {4 \hat \Gamma T \over
\pi \hat U^2 \beta}, \qquad {\rm for}\qquad \hat U \gg 1.
\end{equation}
Diagrammatically, the main contributions to $\hat \chi_\beta(U)$ for systems
with orthogonal symmetry ($\beta=1$) comes from Cooperon and diffuson modes of
diffusion, while for systems with unitary symmetry ($\beta=2$) the Cooperon
degrees of freedom are destroyed by the breaking of T-symmetry. Therefore, when
we croosover from the orthogonal to the unitary ensemble there is a suppresion
of correlations in $\hat \chi_\beta(U)$ by a universal factor 2.

(II) - This case is very interesting since it corresponds to a regime where
perturbation theory and semiclassical approach break down. Using our exact
result, Eq. (\ref{mainresult}), for $\chi_\beta(U)$ we find
\begin{equation}
\chi_\beta(U)=2T\biggl[1-{\beta  \pi^2 \over 2 \gamma }\left({U \over
U_c}\right)^2 h_\beta(\gamma)+O((U/U_c)^4)\biggr],
\label {chi1}
\end{equation}
in which $\gamma=\pi \Gamma /\Delta$ and
$$
h_1(\gamma)=h_2(\gamma)+{1 \over \gamma}+
\int_\gamma^\infty {e^{-t} \over t} dt\left( {d \over d\gamma}
{\sinh \gamma \over \gamma} -{\gamma \over 3}e^{-\gamma}\right),
$$
where
$$
h_2(\gamma)=1+{1-e^{-2\gamma} \over 2 \gamma^2}.
$$
The physical meaning of the first term in Eq. (\ref{chi1}) becomes more
transparent if we consider its application to ballistic mesoscopic
nanostructures and use the definition of $T$ and the Landauer formula for the
average conductance to write
\begin{equation}
\bar G_\beta={e^2 \over h} \chi_\beta(0)=2{e^2 \over h} T=2{e^2 \over h}
{T^R T^L \over T^R + T^L}.
\label{conductance}
\end{equation}
This expression coincides with the average conductance of a quantum dot weakly
coupled to external leads obtained in Ref.\ \onlinecite{prig} using a different
method. It demonstrates that transport in this regime is completely dominated
by tunneling at the junctions and therefore the average Landauer conductance is
just the series addition of the contact conductance associated with the
couplings between the quantum dot and the bulk leads. Note that $\bar G_\beta$
is independent of the size $L$ of the dot in sharp contrast with Ohm's law,
where the diffusive process inside the sample dominates and $\bar G_\beta$
decays linearly with $L$.
It is interesting to observe how $\bar G_\beta$ changes when the coupling to
the leads is strengthened driving the system to the strongly absorbing regime.
In this case the decay width $\Gamma_a$ at each channel becomes comparable to
the mean level spacing and thus $T^R=T^L\simeq \Lambda$ (in the symmetric
case). Note that this regime corresponds to the limit of large number of open
channels and thus it can be treated by both conventional perturbation theory
and semiclassical analysis. For systems with unitary symmetry the total
transmission coefficient becomes equal to the total reflection coefficient
$T=R=\Lambda/2$, as a result of classical ergotic exploration of the boudaries
of the cavity and Eq. (\ref{conductance}) becomes $\bar G_2={e^2 \over h}
\Lambda$. As discussed in Ref.\ \onlinecite{bar} the presence of quantum
interference in systems with orthogonal symmetry leads to a small correction
due to weak-localization and we find $\bar G_1=\bar G_2 +\delta G$.

Finally, the second term in Eq. (\ref{chi1}) determines the curvature
$K_\beta(\gamma)$ of $\chi_\beta(U)$ at $U=0$. The function $h_1(\gamma)$ and
$h_2(\gamma)$ arise ultimately from level repulsion and eigenvector rotations
induced by the external perturbation. One can verify by direct calculation that
$K_1(\gamma) > K_2(\gamma)$ for all $\gamma > 0$.

In the theory of Simons, Altshuler, Lee and Szafer \cite{sim1,sim2,sim3} of
closed chaotic systems in the presence of an external pertubation a remarkable
web of relations has been found \cite{sim3} between the non-linear
$\sigma$-model of diffusion and many other problems in theoretical physics,
such as continuous matrix models, Dyson's Brownian motion model, Sutherland
quantum Hamiltonian and Pechukas gas. In the light of our result, we believe
that similar relations exist for open systems.
An interesting line of research, which we leave for the future, would be to
build a Brownian motion model (or equivalently a quantum Hamiltonian) whose
solution would describe the statistics of the elements of the S-matrix in the
presence of an external perturbation. An immediate application of such
description would be to study parametric fluctuations of the cross-section on
time scales of the order of the Heisenberg time which is well beyond the regime
of validity of semiclassical analysis.

In conclusion, we have studied the effects of an adiabatic external
perturbation on the correlations between different elements of the S-matrix
describing scattering in an open quantum chaotic system weakly coupled to
external reservoirs by two free propagating pipes. We demonstrate that the
parameter dependent autocorrelator of two S-matrix elements  becomes a
universal expression if the perturbation parameter $U$ is measured in units of
the mean square gradient of the energy levels and if the decay width $\Gamma$
is measured in units of the mean level spacing. We have proposed this universal
function as a new signature of quantum chaos in open systems in the weak
coupling regime. We believe that our predictions should in principle be
observable in a microwave-scattering experiment, where both amplitude and phase
of the scattered wave can be accurately measured. The variable $U$ could, for
instance, parametrize changes in the geometry of the chaotic cavity.

The author would like to thank J. T. Chalker for useful comments and
discussions. This work was partially supported by CAPES (Brazilian agency).

\end{document}